\documentclass[amsmath,amssymb, aps, prl, twocolumn]{revtex4-2}
\usepackage{float}
\usepackage{graphicx}
\usepackage{dcolumn}
\usepackage{bm}
\usepackage{hyperref}
\hypersetup{
    colorlinks=true,
    linkcolor=blue,
    filecolor=magenta,      
    urlcolor=cyan,
}
\usepackage{color}
\usepackage{fancyhdr}
\usepackage[normalem]{ulem}

\newcommand{\risp}[1]{\textcolor{black}{{#1}}}
\newcommand{\rizp}[1]{\textcolor{black}{{#1}}}
\newcommand{\rimp}[1]{\textcolor{black}{{#1}}}

\begin{document}

\title{Quantum Zeno Engines and Heat Pumps}

\author{Giovanni Barontini}
\email{g.barontini@bham.ac.uk}
\affiliation{School Of Physics and Astronomy, University of Birmingham, Edgbaston, Birmingham, B15 2TT, UK}

\date{\today}

\begin{abstract}
We study the implementation of quantum engines and quantum heat pumps where the quantum adiabatic transformations are replaced by quantum Zeno strokes. During these strokes, frequent measurements are selectively performed on the external state of the system avoiding transition between different levels. This effectively delivers almost ideal isentropic transformations. We concentrate on the characterization of the performance of a quantum Zeno heat pump implemented with a quantum harmonic oscillator, showing that optimal performance can be achieved faster than with shortcut-to-adiabaticity techniques.   
\end{abstract}

\maketitle

The development of quantum thermodynamics \cite{anders, Goold_2016} is currently providing us with a dynamic and promising framework for the study of quantum processes from the stochastic thermodynamics perspective, for the realization of nano- and micro-engines that take advantage of quantum resources, and for evaluating the energy cost of quantum technologies \cite{prxenergy}. Concerning the development of quantum engines, experimental realizations have demonstrated for example that work can be extracted from a single atom \cite{sciencesinger}, or a single spin \cite{PhysRevLettspin.123.240601}, and that quantum statistics can be used to fuel thermodynamic cycles \cite{widera}. In addition, a plethora of different schemes have been proposed to take advantage of quantum features. Among them, it has been demonstrated that a coherent working fluid could generate more power than an analogous classical one \cite{PhysRevXcoherent.5.031044}, that work could be extracted from quantum correlations with erasure protocols \cite{rio}, that using non-thermal or non-classical baths could lead to more efficient and more powerful engines \cite{scully, PhysRevLettnano.112.030602}, and that a working fluid of interacting quantum particles could be more efficient and more powerful \cite{Jaramillo_2016}. Due to the lack of a classical analogous, quantum measurements have also drawn a particular interest as a resource for the realization of non-trivial quantum engines \cite{PhysRevLettxx.120.260601,PhysRevLettyy.122.070603,PhysRevLettzz.118.260603,PRXQuantumxx.2.040328,PhysRevE.98.042122}. 

\begin{figure}
\centering
\includegraphics[width=0.48\textwidth]{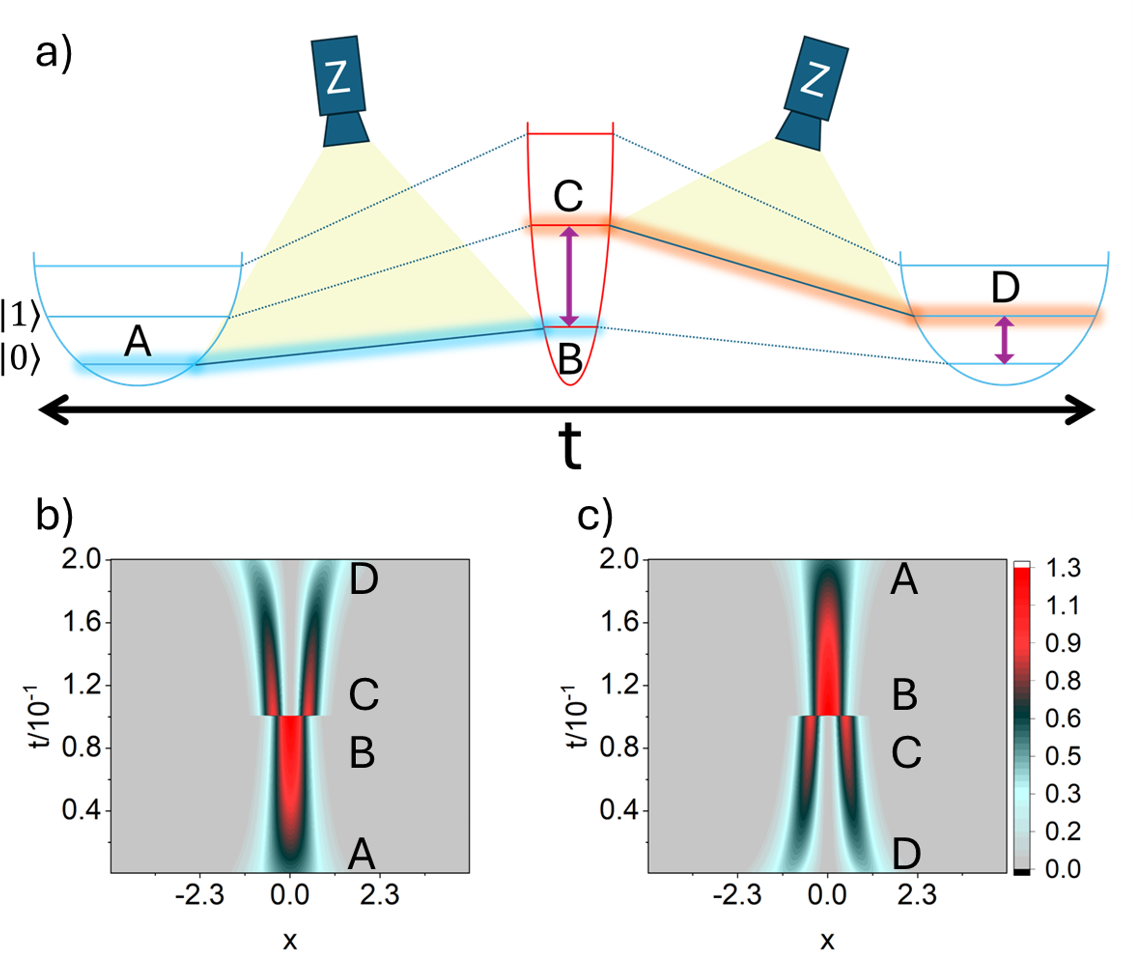}
\caption{a) Schematic representation of a QZE for a one-dimensional harmonic confining potential, \rizp{based on a quantum Otto cycle}. The working fluid is initially prepared in the ground state $|0\rangle$ of the potential. During the A$\rightarrow$B compression a continuous measurement of the $|0\rangle$ state is applied, preventing the working fluid to leave such a state due to the quantum Zeno effect. During the \rizp{isochoric} B$\rightarrow$C stroke the working fluid absorbs heat from a hot bath transitioning to the $|1\rangle$ state. During the following C$\rightarrow$D expansion a continuous measurement of the $|1\rangle$ state is applied, keeping the working fluid in such a state. Finally the thermodynamic cycle is closed coupling the working fluid to a cold thermal bath on which heat is released. If the cycle is reversed, a heat pump is realised. b) Probability density distribution for a working fluid undergoing the protocol shown in a) with $K=5$, $T=10^{-1}$ and $\Omega T=2500$. The B$\rightarrow$C and D$\rightarrow$A strokes have been implemented by applying the $a$ and $a^\dagger$ ladder operators respectively. c) Inverse QZE, or heat pump with the same parameters as b).}
\label{Fig1}
\end{figure} 

In this Letter, we propose to utilize the quantum Zeno effect to substantially boost the power of quantum engines and quantum heat pumps. Adiabatic (or isentropic) transformations are needed in the vast majority of quantum thermodynamic cycles. These are usually the limiting factor for generating high power outputs because they intrinsically require long times to minimise dissipation. Considering a quantized working fluid undergoing a compression or decompression of its energy levels, the work done by or on the system can be written as $W=\sum_nP_ndE_n $, with $E_n$ and $P_n$ the energy and occupation probability of the $n$-th level \cite{nori}. If the transformation is faster than $\simeq \hbar/\Delta E_n$, with $\hbar$ the reduced Planck constant and $\Delta E_n$ the separation between adjacent energy levels, the transition probability between the energy levels becomes non-negligible, and the system experiences dissipation. Notably, advanced techniques such as shortcut-to-adiabaticity \cite{paternostro,Barontini_2019} can achieve ideal isentropic transformations in finite time, substantially improving the power output of quantum or miniaturized engines. Such techniques possess however a cut-off time, that could be surpassed with the use of the quantum Zeno effect. We show here that by continuously and selectively measuring a quantum system undergoing a compression or decompression of its energy levels, it is possible to `freeze' the populations $P_n$ and achieve almost perfect adiabatic transformations in times shorter than those obtained with shortcut-to-adiabaticity techniques. In particular, we concentrate on the characterization of a quantum heat pump where the working fluid is a quantum harmonic oscillator. We show that the usual requirement for achieving adiabaticity $\dot{f}/f^2\ll1$, with $f$ the harmonic oscillator frequency, can be relaxed to $\dot{f}/f^2\ll M$, with $M$ the number of measurements performed during the adiabatic Zeno stroke. This leads to a speeding up of the adiabatic transformations of orders of magnitude, that could directly translate in orders of magnitude higher output power.     

\begin{figure}
\centering
\includegraphics[width=0.48\textwidth]{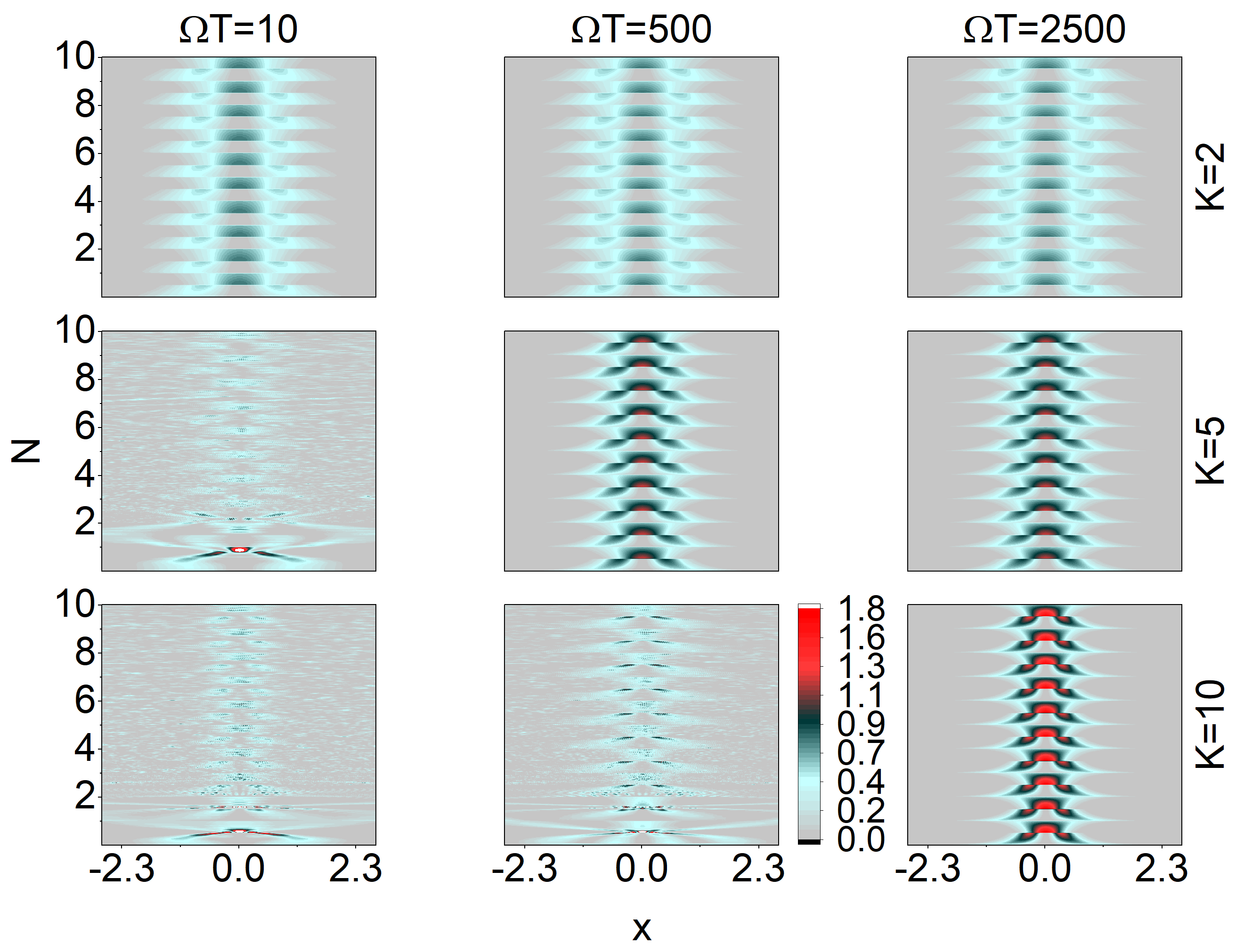}
\caption{Evolution of the probability density distribution of the working fluid undergoing ten cycles of the QZHP, where the duration of each Zeno stroke is $T=5\times10^{-2}$. The rows correspond to different compression ratios $K$, while the columns correspond to different number of measurements $M=\Omega T$ during the Zeno strokes.}
\label{Fig2}
\end{figure} 

We will consider as working fluid an object that has been cooled down to the ground state of a confining harmonic potential $V=m(2\pi fx)^2/2$, with $m$ the mass of the working fluid. In the following, we will express distances in units of the harmonic oscillator length $a_{ho}=\sqrt{\hbar/(2\pi f m)}$, time in units of $f^{-1}$, and energies in units of $2\pi\hbar f$. The working principle of the QZE is shown in Fig. \ref{Fig1} a). The object is initially prepared in the ground state $|0(t=0)\rangle$ of the harmonic potential. During A$\rightarrow$B, work is performed on the object by compressing the potential in a time $T$, reaching the final frequency $f_f=Kf$, with $K$ the compression ratio. During the compression, frequent measurements on the $|0\rangle$ state are performed. This corresponds to projecting the state of the working fluid on the instantaneous ground state of the potential, i.e., to applying the $|0(t)\rangle\langle 0(t)|$ operator, with a rate $\Omega$. If $\Omega T=M$ is sufficiently large, the quantum Zeno effect prevents the object to leave the $|0(t)\rangle$ state at every time, even if $T\ll1$. This effectively realises a perfect, or frictionless, isentropic compression for finite values of $T$. Interestingly, according to \cite{abdelkhalek2018fundamental}, our implementation of the Zeno effect comes at vanishing costs from the projective measurements, \risp{see \cite{SM}}.

During B$\rightarrow$C the working fluid absorbs heat from a hot bath, \rizp{utilizing for example an isochoric transformation as in Fig. 1}, populating the $|1\rangle$ state. During C$\rightarrow$D, the working fluid does work expanding in the potential, that relaxes back to the initial frequency $f$ in a time $T$. During the decompression, the Zeno effect is enforced by applying the $|1(t)\rangle\langle 1(t)|$ operator with frequency $\Omega$, keeping the object in the $|1(t)\rangle$ state at every time. Finally the D$\rightarrow$A \rizp{isochoric} stroke closes the thermodynamic cycle and the working fluid releases heat to a cold bath. In Fig. \ref{Fig1} b) we report the evolution of the probability density distribution of the working fluid during the QZE. Note that despite the total cycle time being only $\simeq0.2$, the working fluid remains in the ground state $|0\rangle$ during the A$\rightarrow$B compression, and in the first excited state $|1\rangle$ during the C$\rightarrow$D expansion.  
If the cycle is reversed a heat pump or refrigerator is realised. Fig. \ref{Fig1} c) shows the evolution of the probability density distribution of the working fluid during an inverse QZE with the same parameters as Fig. \ref{Fig1} b). Also in this case there is no apparent sign of dissipation.

Several schemes have been proposed to realise quantum engines \rizp{based for example on quantum Carnot, Otto, Diesel and Braydon cycles}, \risp{see e.g. \cite{annurev:/content/journals/10.1146/annurev-physchem-040513-103724, anders, nori, Mukherjee_2021, fava2, favaPhysRevResearch.2.043247}}, and the Zeno strokes could in principle be used to replace the isentropic strokes in all of them, if a suitable experimental platform exists. The details of the B$\leftrightarrow$C and D$\leftrightarrow$A strokes will not be discussed in this work, as these do not differ from other kinds of engines. The configuration that we will mostly consider is the one in which the inverse QZE \rizp{utilizing an inverse Otto cycle} acts as a heat pump for electromagnetic fields, as in Fig. \ref{Fig1} c). \rizp{A characterization of a QZE utilising a Otto cycle as the one shown in Fig. 1 b) is instead provided in \cite{SM}.} The quantum Zeno heat pump (QZHP) can be used to `transform' a photon to another one with $K-1$ more energy. During the A$\rightarrow$D stroke the working fluid absorbs a photon of energy 1 and undergoes a population swap by means of a $\pi$ pulse of resonant radiation. 
During the C$\rightarrow$B stroke the working fluid emits a photon of energy $K$ applying another $\pi$ pulse. 
Because $\pi$ pulses can be controlled with very high degree of precision and can in principle be arbitrarily short, we will describe these strokes with the instantaneous application of the ladder operators $a^\dagger$ and $a$ respectively at the end of the Zeno strokes, as done in Fig. \ref{Fig1} b) and c). In the conclusions we will discuss a possible technological platform enabling the practical implementation of the quantum Zeno engines and heat pumps. 

\begin{figure}
\centering
\includegraphics[width=0.49\textwidth]{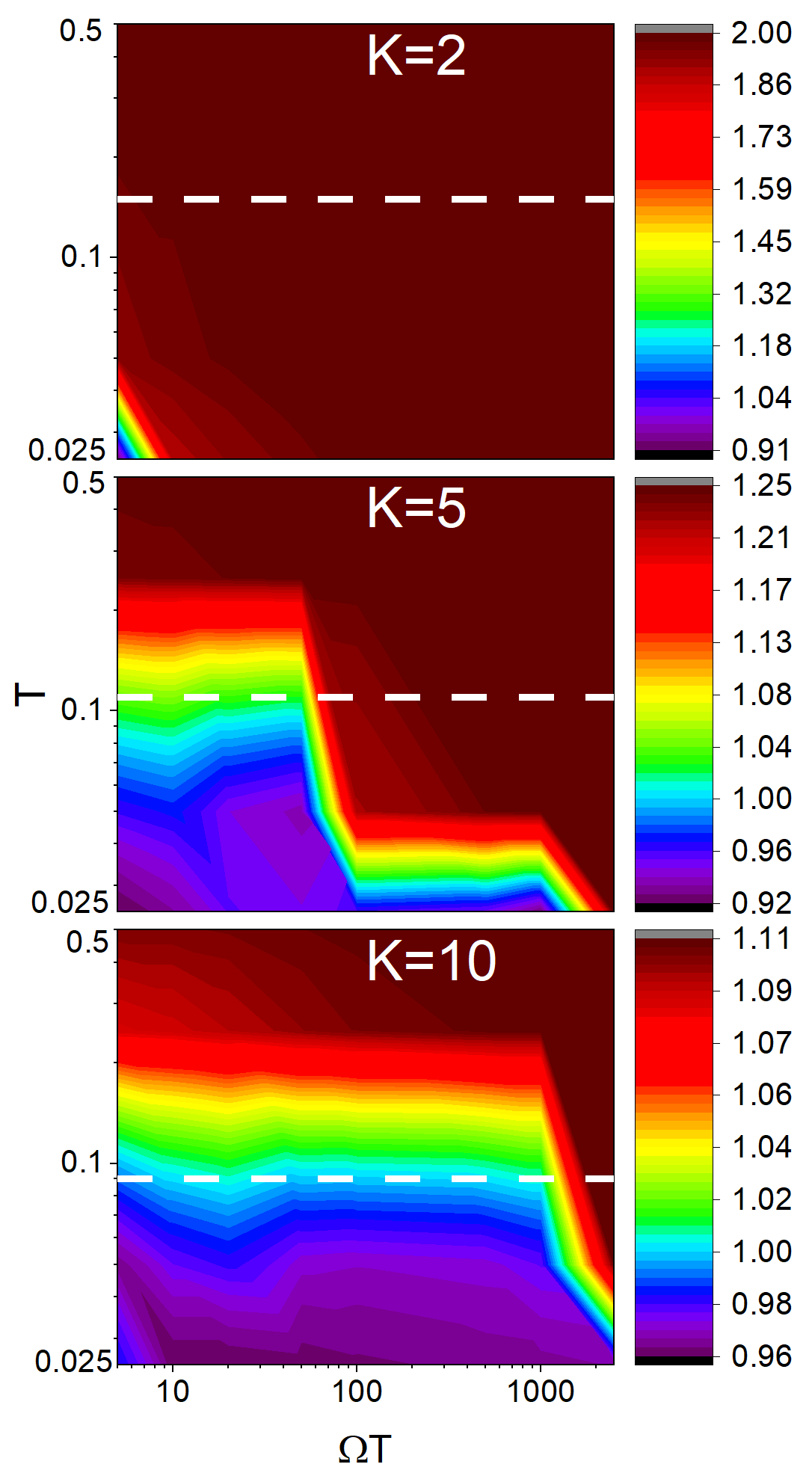}
\caption{Average value of the coefficient of performance $\bar{\xi}$ over 100 cycles of the QZHP \rizp{utilizing an inverse Otto cycle} for three different values of the compression ratio $K$ as a function of the number of measurements $\Omega T$ and the duration of the Zeno strokes $T$. For $K=2$ $\xi_{opt}=2$, for $K=5$ $\xi_{opt}=1.25$ and for $K=10$ $\xi_{opt}=1.11$. Tho horizontal dashed lines correspond to the cut-off time for shortcut-to-adiabaticity protocols \cite{paternostro}, \rizp{i.e. it is not possible to implement shortcut-to-adiabaticity protocols for lower values of $T$}.}
\label{Fig3}
\end{figure} 

To provide a concrete example, we numerically simulate the one-dimensional QZHP \rizp{based on the inverse Otto cycle depicted in Fig. 1}, with split-step methods on a grid of 512 sites and total length $\simeq$9.3. The time step used is $10^{-5}$. As discussed, the A$\rightarrow$D stroke consists on the application of the $a^\dagger=(\hat{x}-i\hat{p})/\sqrt{2}$ operator, while the C$\rightarrow$B stroke on the application of the $a=(\sqrt{K}\hat{x}+i\hat{p}/\sqrt{K})/\sqrt{2}$ operator. During the D$\rightarrow$C Zeno stroke, $f$ is linearly ramped to $Kf$ in a time $T$ and the $|1(t)\rangle\langle 1(t)|$ operator is applied every $\tau=1/\Omega$, where $|1(t)\rangle$ is the first excited state of the harmonic oscillator potential $V(t) = \{[1+(K-1)t/T]x\}^2/2$. the B$\rightarrow$A Zeno stroke is implemented in the same way, with the harmonic frequency linearly ramped down from $Kf$ to $f$, and applying the $|0(t)\rangle\langle 0(t)|$ operator.  
In Fig. \ref{Fig2} we show the evolution of the probability density distribution of the working fluid undergoing $N=$10 cycles in which each Zeno stroke lasts $T=5\times10^{-2}$, for different values of $K$ and $\Omega$. For low compression ratios, the Zeno strokes are isentropic with good approximation even with sparse measurements. For $K=2$, the working fluid goes through the series of compressions and decompressions without any sign of dissipation even if only 10 measurements are performed during $T$, as shown in the first row of Fig. \ref{Fig2}. As $K$ increases however, it becomes harder to avoid dissipation during the Zeno strokes, note that increasing $f$ by a factor of $K$ corresponds to increasing $V$ by a factor $K^2$. For $K=5$ one needs to perform at least 100 projective measurements to avoid dissipation. For lower values of $\Omega$ the target states are populated only with fractional probability. According to \cite{abdelkhalek2018fundamental}, in this case the cost of the Zeno strokes diverges. For $K=$10, the substantially higher value of $\Omega=2500$ is needed to keep the QZHP running with negligible dissipation.

To quantitatively evaluate the performance of the QZHP, we calculate its coefficient of performance, defined as $\xi=Q_{out}/(Q_{out}+Q_{in})$, with $Q_{in}>0$ and $Q_{out}<0$ the heat absorbed and emitted respectively. For a quantum harmonic oscillator we can define the heat exchanged as $Q=\sum_nE_ndP_n=\sum_nE_n[P_n(X)-P_n(Y)]$, with $X=C,A$ and $Y=B,D$ \cite{nori}. The optimal value of $\xi$ is therefore $\xi_{opt}=K/(K-1)$, that is achieved when the probability of finding the working fluid in the state $|1\rangle$ ($|0\rangle$) is 1 at the end of the Zeno compression D$\rightarrow$C (decompression B$\rightarrow$A). Note that $\xi_{opt}$ decreases as $K$ increases. In Fig. \ref{Fig3} we report the average value of the coefficient of performance $\bar{\xi}$ over 100 cycles of the QZHP for different compression ratios as a function of the number of measurements $M=\Omega T$ and of the duration of the Zeno strokes $T$. As already suggested by the behaviour of the probability density distribution, the QZHP is able to reach $\xi_{opt}$ also when the Zeno strokes last for $T<1$, provided that a sufficient number of measurements is performed to enable the Zeno effect. For modest compression ratios the duration of the Zeno strokes can be lower than $10^{-2}$ and still achieving $\bar{\xi}=\xi_{opt}$. Notably, for all compression ratios, it is possible to find values of $\Omega T$ that enable the Zeno strokes to be faster than shortcut-to-adiabaticity protocols, as shown in Fig. 3. \rizp{The same conclusions can be drawn evaluating the amount of entropy produced during the Zeno strokes, see \cite{SM}.} \risp{Considering for example a harmonic oscillator with $f=10$ kHz, which is a typical value for cold atoms in optical tweezers, one could achieve $\xi\equiv\xi_{opt}$ with Zeno strokes of $T\simeq$2.5 $\mu$s for $K=2$ and 5, and $T\simeq$5 $\mu$s for $K=10$.} \rizp{In \cite{SM} we provide a similar characterization for the efficiency of the QZE.}

\begin{figure}
\centering
\includegraphics[width=0.48\textwidth]{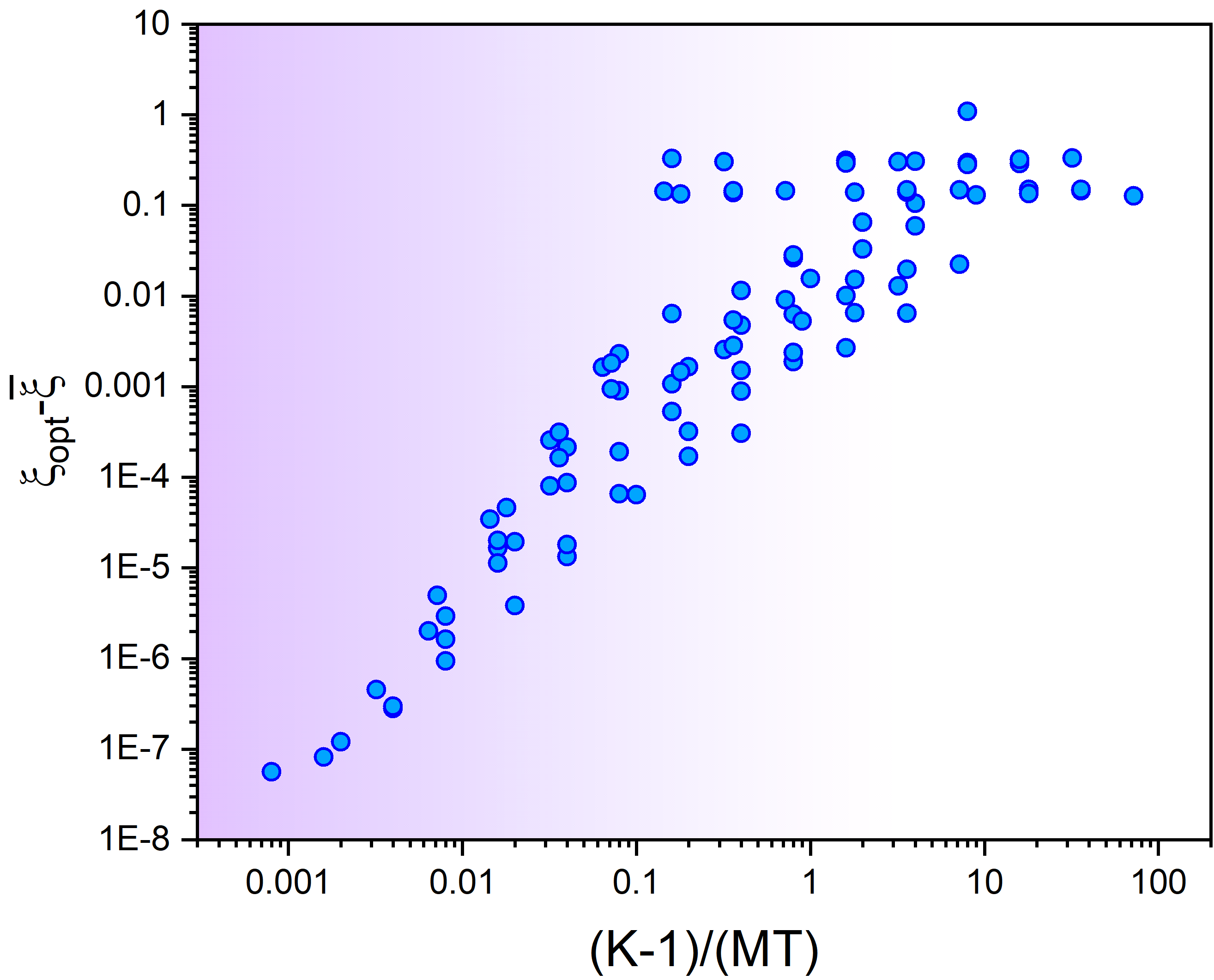}
\caption{Difference between the average value of the coefficient of performance $\bar{\xi}$ over 100 cycles of the QZHP and its expected optimal value $\xi_{opt}$ as a function of the parameter $(K-1)/(MT)\equiv\dot{f}/M$ for the data reported in Fig. \ref{Fig3}.}
\label{Fig4}
\end{figure}

Insight on the observed behaviour can be gained by looking at the evolution of the populations $P_n(t)=|c_n(t)|^2=|\langle n(t)|n(t)\rangle|^2$ during the Zeno strokes, where $|n(t)\rangle$ are the instantaneous eigenstates of the time-varying Hamiltonian, defined as $H(t)|n(t)\rangle=E_n(t)|n(t)\rangle$. If we consider for example a Zeno stroke where $P_k(t=0)=1$, it is straightforward to show that the evolution of the populations at short times in one of the other states is given by $P_{n\neq k}(t)\simeq (\dot{H}_{kn}/\hbar\omega_{kn})^2t^2$, where $\dot{H}_{kn}(t)=\langle k(t)|\dot{H}(t)|n(t)\rangle$ and $\omega_{kn}(t)=[E_n(t)-E_k(t)]/\hbar$. In our case the probability of populating not neighbouring states decrease quadratically, so it can be ignored, and we obtain $P_{k}(t)\simeq 1-\langle k(t)|\{x^2[1+(K-1)t/T](K-1)/T\}^2|k\pm1(t)\rangle t^2$, that is 1 for $(K-1)/T=\dot{f}\ll1$, i.e., for very slow or very little frequency variations. If we now perform frequent measurements of the state $|k(t)\rangle$, the survival probability after $M=\Omega T$ measurements in a time $T$ is $P^{(M)}_k(T)=P_k(T/M)^M\simeq \{1-\langle k(t)|\{x^2[1+(K-1)/M](K-1)/T\}^2|k\pm1(t)\rangle (T/M)^2\}^M$ \cite{Facchi_2009}, that remains $\simeq1$ for $(K-1)/T\ll M$, therefore enabling a speeding up of the frequency change of a factor $M$. \risp{Note that the use of instantaneous eigenvalues to contruct the projector operators is well justified since the projection operation itself is instantaneous. In our simulations the application of the projective operator lasts 10$^{-5}$, given by the time step, during which the relative change in $f$ is negligible.} In Fig. \ref{Fig4} we report the difference between $\bar{\xi}$ and $\xi_{opt}$ for all the data shown in Fig. \ref{Fig3} as a function of $(K-1)/(MT)$. In accordance with the discussion above, for values of $(K-1)/(MT)$ lower than 0.1, the average discrepancy between the coefficient of performance of the QZHP and the expected optimal value is below $10^{-3}$. 

In summary, we have shown that the performance of quantum engines and quantum heat pumps can be substantially improved if quantum adiabatic transformations are replaced with quantum Zeno strokes. Quantum adiabatic transformations are usually the bottleneck preventing high power quantum engines and heat pumps, as they require long times to avoid dissipation. The quantum Zeno effect enables to perform the same amount of work on and by the working fluid in a finite time with negligible dissipation, also surpassing shortcut-to-adiabaticity techniques. We have studied a system composed by a working fluid confined within a quantum harmonic oscillator, and we have used it to construct a quantum heat pump \rizp{and a quantum engine \cite{SM} based on the Otto cycle} where the adiabatic transformations are replaced by quantum Zeno strokes. We have shown that such a QZHPs and QZEs can achieve optimal performance even when the duration of the quantum Zeno strokes is substantially lower than 1. We have shown that the speed of the quantum adiabatic transformations can be increased by a factor of $M=\Omega T$, that directly translates in faster quantum thermodynamic cycles. 

The practical implementation of QZEs and QZHPs requires the ability to perform quantum non-demolition measurements, i.e., that during the Zeno strokes no energy is exchanged between the measurement device and the working fluid. Another strict requirement is the selectivity of the measurement, that needs to neatly separate the state under measurement with the rest of the Hilbert space. Possible experimental platforms include quantum dots \cite{qdot} and trapped ions \cite{steiner2013single} or neutral atoms \cite{volz2011measurement} in optical cavities. \rimp{From \cite{abdelkhalek2018fundamental}, one would conclude that the energy cost associated to the Zeno strokes proposed here corresponds only to the vanishing amount of entropy produced \cite{SM}, apparently achieving lower costs than shortcut-to-adiabaticity methods \cite{qqqPhysRevResearch.2.023120,wwwPhysRevE.99.022110,eeePhysRevA.94.042132}. However, the impact of experimental imperfections such as non-perfect non-demolition measurement would need to be carefully evaluated for each experimental platform, because any exchange of energy during the projective measurement would lead to an exponentially increasing cost of the Zeno strokes.} To give an example, a possibility to experimentally realise QZEs and QZHPs is to utilise a clock transition to interrogate an atom trapped in a harmonic potential inside a high finesse ring cavity \cite{popplau}. This will ensure that both the linewidth of the transition addressed and of the cavity are $\ll f$, and that the probe field does not exhibit spatial dependence. The $\pi$ pulses corresponding to the B$\leftrightarrow$C and D$\leftrightarrow$A strokes could instead be realized for example with a two-photon Raman configuration tuned on the blue or red sideband respectively. \risp{An intriguing possibility is to utilize Zeno strokes in single \cite{PhysRevLettspin.123.240601} and many-body \cite{Mukherjee_2021, fava2, favaPhysRevResearch.2.043247} spin systems, that are promising towards the implementation of quantum engines and quantum batteries \cite{fava1}. For example, in engines based on transverse Ising systems, see e.g. \cite{fava2,favaPhysRevResearch.2.043247}, one could implement the Zeno strokes by performing frequent measurements on the number of excitations in the system, provided that the tranverse field is maintained far away from the quantum critical point. For more complex many-body systems, the main hurdle is represented by the fact that the implementation of the Zeno strokes requires the knowledge of the instantaneous eigenstates, which is often rather challenging.}

In conclusion, this work shows that the quantum Zeno effect could be an effective tool to be added to the quantum toolbox for the enhancement of the performance of quantum engines and heat pumps towards their practical implementation in quantum information science. 

\paragraph*{Acknowledgements}
I acknowledge fruitful discussions with M. Paternostro, A. Balenchia, R. Puebla, and S. Gherardini. I thank V. Guarrera and A. Deb for reading the manuscript and useful comments. acknowledge the use of computing power provided by the Advanced Research Computing centre at the University of Birmingham.

\bibliography{main_bibl}

\end{document}